\documentclass[english,prb,aps,showpacs]{revtex4}
\usepackage{amsmath}
\usepackage{amssymb}
\usepackage{graphicx}
\usepackage{epsfig}
\usepackage{amsfonts}

\begin{document}

\title{Modification of the Bloch law in ferromagnetic nanostructures}
\author{S. Cojocaru$^{1}$,  A. Naddeo$^{2}$ and R. Citro$^{2,3}$}
\affiliation{$^{1}$Horia Hulubei National Institute for Physics and Nuclear Engineering,
P.O.B. MG-6, 077125 Magurele, Romania;\\
$^{2}$Dipartimento di Fisica \textquotedblleft
  E.R.Caianiello\textquotedblright , Universit\`{a} degli Studi di Salerno,
  84084 Fisciano (Sa), Italy;\\
$^{3}$Spin-CNR, Unit of Salerno, 84084 Fisciano (Sa), Italy.}

\begin{abstract}
The temperature dependence of magnetization in ferromagnetic nanostructures (e.g., nanoparticles or nanoclusters) is usually analyzed by means of an empirical extension of the Bloch law sufficiently flexible for a good fitting to the observed data and indicates a strong softening of magnetic coupling compared to the bulk material. We analytically derive a microscopic generalization of the Bloch law for the Heisenberg spin model which takes into account the effects of size, shape and various surface boundary conditions. The result establishes explicit connection to the microscopic parameters and differs significantly from the existing description. In particular, we show with a specific example that the latter may be misleading and grossly overestimates magnetic softening in nanoparticles. It becomes clear why the usual $T^{3/2}$ dependence appears to be valid in some nanostructures, while large deviations are a general rule. We demonstrate that combination of geometrical characteristics and coupling to environment can be used to efficiently control magnetization and, in particular, to reach a magnetization higher than in the bulk material.
\end{abstract}
\pacs{75.75.-c,73.21.-b,75.50.Tt}
\maketitle

\section{Introduction}

It is well known that size, shape and surface effects play an important role in nanostructures, e.g. nanoparticles and nanoclusters (NP and NC)
 \cite{Guimaraes}. For a macroscopic crystal these effects can be neglected because its surface to volume ratio is sufficiently small and despite of a formally broken translation symmetry, one can still use periodic boundary conditions to accurately describe the observed bulk behavior. A typical example is the famous Bloch $T^{3/2}$ law for the temperature dependence of magnetization in ferromagnets
\cite{Kittel}. However, even in the \textquotedblleft macroscopic limit\textquotedblright, this picture is incomplete as demonstrated by formation of magnetic domain walls as a result of energy balance between the demagnetization (or stray) field and the magnetocrystalline anisotropy. For sufficiently small (dozens of nm) samples a single domain structure is stabilized, resulting in a large uncompensated magnetic moment $\vec{\mu}\left( T\right) V$ (we use lattice spacings as length units so that $V$ is the number of atoms), $\mu =\left\vert \vec{\mu}\right\vert $ is the magnitude of internal magnetization \cite{Apsel}.
The latter is directly related to the superparamagnetic relaxation at temperatures above $T_{B}$ ( $T_{B}<T<T_{C}$, where $T_{C}$ is the Curie
temperature) and to the blocking effect at lower temperatures. The magnetocrystalline anisotropy field causing the blocking effect is $H_{A}\thicksim 2K/\mu $, see e.g., \cite{morup}, where $K$ is the anisotropy constant. We further consider the case of small $K$ as it may be several orders smaller than the exchange coupling $J$, the dominant interaction for nanoparticles below $40-50$ nm in size  \cite{zhang}.  Typically $H_{A}$ is a fraction of $Tesla$,
e.g., in a $Fe$ bcc crystal, $K$ is about $2.4$ $\mu eV/$atom, while $J$ reaches dozens of $meV$. In the superparamagnetic regime ( $T > T_{B}$ ) the magnetization vector $\vec{\mu}$  fluctuates between the easy directions and the time average of its projection  $\mu _{eff} $   on the applied field  $H_{e}$  is described by the Langevin function
\begin{equation}
\mu _{eff}=\mu \mathcal{L}\left( \beta \mu VH\right) =\mu \left[ \coth
\left( \beta \mu VH\right) -\frac{1}{\beta \mu VH}\right],  \label{1}
\end{equation}%
where $\beta=T^{-1} $ ($k_{B}=1$). For this temperature interval Eq. (\ref{1}) actually serves to
determine $\mu $ from the measured value of $\mu _{eff}$ \cite{Apsel}.  For a sufficiently large applied field $\mu $ is identified as saturation magnetization. However, the
temperature dependence $\mu =\mu\left( T\right)$ is still poorly
understood and remains an open topic of experimental and theoretical research. For instance, by decreasing the size $V$ one usually observes
large deviations of $\mu\left( T\right)$ from the Bloch law and an unpredictable variation of
its parameters. In this context more puzzling is the standard $T^{3/2}$
behavior reported for some NP \cite{Markovich}.

When $T<T_{B}$   the vector  $\vec{\mu}$   precesses at an angle $\theta$ near the anisotropy axis, so that although  $\theta$  is subject to thermal fluctuations its projection $\mu _{eff}\neq 0$ even if $H_{e}$ vanishes (blocking effect) . In this case Eq. (1) is replaced by  \cite{morup}
\begin{equation}
\mu _{eff}\left( T\right) =\mu \left( 0\right) \left( 1-T/2KV\right),
\label{3a}
\end{equation}%
i.e., the $T-$ variation of the internal magnetization is neglected. On the other hand, numerical diagonalization study of ferromagnetic clusters in a series of seminal papers \cite{Hend,Hendriksen} has led to the phenomenological expression for the internal magnetization
 $T\lesssim 0.5T_{C} $:
\begin{equation}
\mu _{phen}\left( T\right) =\mu \left( 0\right) \left( 1-\gamma T^{\ \alpha
}\right) .  \label{2a}
\end{equation}%
Eq. (\ref{2a}) is sufficiently flexible to reproduce the temperature dependence and is
currently widely used for the description of experimental
and numerical simulation data on $\mu\left( T\right)$, e.g.  \cite{ortega, gubin}.
In the macroscopic limit, $V\rightarrow \infty$ , one recovers the Bloch result with $\alpha =3/2$ and $\gamma \sim D^{-3/2},$ where $D$ is the stiffness of magnon spectrum ( $D \sim J$, and $J$ is the exchange coupling constant). However, the phenomenological parameters $\alpha $ and $\gamma $ obtained in this way do not show a regular variation with size, e.g. $\alpha $ may be found scattered between $1.5$ and $3$ (see, e.g., Ref.\cite{Hend, Demortier}) and
the reason for such a non-monotonous behavior is unclear. It is also known that NP magnetization depends on its geometric shape and is
affected by contact with an external medium, e.g. by embedding or coating. These properties are of major importance for applications and, at the same time, raise a fundamental challenge for understanding their connection to the shape and size of a nanostructure or to the specific boundary
conditions on its surface. From the phenomenological description it is not possible to tell how this information is encoded in the fitting parameters or whether these can be given a microscopic meaning at all.

In the following the above issues are discussed in the context of the
proposed generalization of the microscopic Bloch theory of bulk ferromagnets. An important new aspect is that, in contrast to the bulk material, coupling to environment can affect not just the surface of a nanostructure but its properties as a whole. An essential role is played by the collective excitations (magnons), the mediators of such effects. On the one side, magnons are sensitive to  boundary conditions (BC) and at the same time involve all the constituent magnetic moments of the crystal; on the other side, they dominate the low energy part of the magnetic excitation spectrum. These features are well captured by the quantum spin-$S$ Heisenberg model with nearest-neighbour exchange coupling. The shape anisotropy is taken into account by considering a rectangular crystal of a general form $N_{x}\times N_{y}\times N_{z}=V$, while the effect of environment is modeled by the choice of boundary conditions on its surface. It is clear that standard periodic BC (PB) are even qualitatively insufficient since existence of a surface is totally ignored in this case. We therefore use a $3D$ generalization of the Heisenberg spin chain
Hamiltonian considered in \cite{rrp}, where the BC are controlled by the
external field $\nu $ acting on the crystal surface. As will be shown below
a realistic description is achieved by analysing the main classes of possible BC
at finite values of the coupling field. Some physical implications of the results will be demonstrated with a specific example of quantum Monte-Carlo cluster simulation.

\section{Finite temperature magnetization}

In a finite system it is important to make a distinction between the
uniform excitation mode \cite{morup}, $\mathbf{q}=0$ \textquotedblleft magnon \textquotedblright,
and the \textquotedblleft usual\textquotedblright or dispersive magnon modes
because they contribute to the observed $\mu _{eff}$ in a
different way. The latter are mainly responsible for the temperature
dependence of the magnitude of the magnetization $\mu \left(
T\right) ,$ while the former accounts for the thermal fluctuations of its
direction (which defines the Z - axis) with respect to the anisotropy axis, i.e. the $\mathbf{q}=0$ mode describes the rotation of the vacuum of the dispersive magnons. This can be understood by
examining the magnon dispersion for a simple cubic lattice :
$\varepsilon _{\mathbf{q}}=2JS\sum_{\xi \ =x,y,z}\left( 1-\cos q_{\xi
}\right) +g\mu _{B}H_{A}, $
where $\mu _{B}$ is the Bohr magneton and Planck's constant is omitted. The
spectrum of a finite magnet is gapped and at low energies can be approximated as $\varepsilon _{\mathbf{q}}\sim J\sum_{\xi}\left(
m_{\xi}/N_{\xi}\right) ^{2}+g\mu _{B}H_{A},$ where $m_{\xi}=0,1,2,...$\ . For not too large $N_{\xi}$ the
energy scale $\omega _{A}=g\mu _{B}H_{A}$ of the uniform ($m_{\xi}=0$)
mode is lower than the magnon excitation gap between $m_{\xi}=0$ and
$m_{\xi}=1$. When $T<T_{B}$ the temperature dependence of $\mu _{eff}$ appears to be dominated by the uniform mode described by harmonic
oscillations of the magnetization axis  ($Z-$ axis) in the potential well created by magnetocrystalline anisotropy and, as already mentioned, is described by Eq. (\ref{3a}). This expression is well justified when thermal energy is lower than the
magnon gap, $T\lesssim JS\pi ^{2}/N^{2}\equiv T^{\ast },$ since then $\mu
\left( T\right) $ is exponentially close to $\mu \left(
0\right) $. However, the linear temperature dependence of $\mu _{eff}$ for the interval $T^{\ast
}<T<T_{B}$  was deduced from the assumption that $\mu \left(
T\right)$ follows the Bloch law or its phenomenological generalization in Eq.  (\ref{2a}), where $%
\alpha >3/2$, see \cite{Hend}. Instead, the microscopic analysis presented below shows
that $\mu \left( T\right) $ contains quasi-linear $T -$ components and therefore its temperature dependence can not be ignored.

To calculate the low temperature expansion of $\mu \left(T\right) $ we neglect the anisotropy field in magnon
dispersion (its inclusion will not qualitatively change the
results) and exponentially small terms like $\exp \left( -\beta J\right) :$
\begin{equation}
\Delta \mu \equiv 1-\mu \left( T\right) /\mu \left( 0\right) =\frac{1}{VS}%
\sum_{n=1}^{\infty }\sum_{\mathbf{q}\neq 0}\exp \left( -\beta n\varepsilon _{%
\mathbf{q}}\right) .  \label{4}
\end{equation}%
The issue is whether there exist a general formula that embodies the diversity of physical conditions defined by size, shape and BC contained in Eq.(\ref
{4}). The problem of surface boundaries in a $3D$ system
presents an additional interest because even the analytic form of the magnon
dispersion has not been discussed before except for limit cases. We
will therefore consider the examples representative of the main classes of
BC. In numerical simulations the most studied is the case of free BC (FB),
when magnon wave numbers in Eq. (\ref{3a}) are quantized as follows:
$q_{\xi }^{FB}=\frac{\pi m_{\xi }}{N_{\xi }};\ \ m_{\xi }=0,..,N_{\xi }-1.$
Clamped BC (CB) represent a limit opposite to FB, when the surface spins are virtually fixed by the strong border fields
$\nu \rightarrow \infty$. This is clearly demonstrated by the form of
the respective magnon wave functions which are coordinate products of $\cos
\left( q_{x}^{FB}\left( X-1/2\right) \right) \times ... \times \cos
\left( q_{z}^{FB}\left( Z-1/2\right) \right) $ and $ \sin
\left( q_{x}^{CB}X\right) \times...\times \sin
\left( q_{z}^{CB}Z\right) $ (compare to the uniform distribution of spin deviations for the periodic BC).
Fortunately, the discrete Schrodinger equation can be solved also at finite values of the interface fields. Namely, we will consider the case of a particle embedded (EB) in a medium with coupling comparable to the internal fields of the NP, i.e. $\nu =JS:$%
\begin{equation}
q_{\xi }^{EB}=\frac{\pi m_{\xi }}{N_{\xi }+1};\ \ \ \xi =x,y,z;\ \ m_{\xi
}=1,..,N_{\xi }.  \label{6}
\end{equation}%
Note that for a different coupling strength quantization of collective
excitations may slightly differ, e.g. for $EB2$ when $\nu =2JS$ we should
replace $N_{\xi }+1\rightarrow N_{\xi }$ in Eq. (\ref{6}), while magnon
amplitudes preserve the $\sin $ -function dependence similar to CB. In
addition to the above uniform BC one may define another
physically distinct class, mixed BC (MB), when the coupling to environment
is non-uniform and may differ from side to side. As a representative we take
the example of a NP with all sides free except one at $Z=0$. For the
same coupling strength as in Eq. (\ref{6}) we obtain the
following wavenumbers:%
\begin{equation}
q_{z}^{MB}=\frac{\pi \left( 2m_{z}+1\right) }{2N_{z}+1};\ \ \
m_{z}=0,..,N_{z}-1.  \label{7}
\end{equation}
These eigenvalues are further used to calculate the quantum
statistical average in Eq. (\ref{4}), as for instance:
\begin{eqnarray}
&& \Delta \mu ^{FB}\left( T\right)  =  \label{4a} \\
&& \sum_{n=1}^{\infty }\frac{1}{SV}\left( \prod\limits_{\xi }\sum_{m_{\xi
}=0}^{N_{\xi }-1}\exp \left[ -4n\beta JS\sin ^{2}\left( \frac{\pi m_{\xi }}{%
2N_{\xi }}\right) \right] -1\right) .  \notag
\end{eqnarray}%

Every finite sum over quantum numbers $\left\{ m_{\xi }\right\} $ defined by
specific BC can be transformed into a series of modified Bessel functions $%
I_{2k_{\xi }N_{\xi }}\left( 2\beta nJS\right) $ ($k_{\xi }=0,1,2,...,\infty
$) by generalizing the formalism in \cite{ijmp}. The size independent term $%
k_{x}=k_{y}=k_{z}=0$ in this expansion corresponds to the macroscopic limit,
i.e. to the Bloch $T^{3/2}$ law, as follows from the standard asymptotic
expression of the Bessel function: $\left( \exp \left( -2\beta nJS\right)
I_{0}\left( 2\beta nJS\right) \right) ^{3}\varpropto \left( 4\pi n\beta
JS\right) ^{-3/2},$ when $T/J\ll 1$. However the finite-$N$ terms cannot be
treated in the same way because in addition to the large argument $2\beta nJS
$ we now have three more (i.e.$k_{\xi }N_{\xi }$) parameters taking large
values in the order index of the Bessel function and the known uniform
asymptotics \cite{olver} does not lead to further progress. This difficulty
can be circumvented if we renounce to the straightforward $T$ expansion and
consider an expansion in the scaled parameters $P_{x}^{FB}=TN_{x}^{2}/\pi JS, etc.$
These parameters represent the ratios between thermal energy and
the lowest magnon gap for a given BC (divided by $\pi $ for computational reasons). The
integral representation of the Bessel function allows then to develop a
generalized low temperature expansion in terms of the $P -$ variables. It is
still a \textquotedblleft low $T$\textquotedblright, but the condition $%
T/J\ll 1$ is now detailed depending on the physical regime controlled by the
parameters $P$. The derivation is further explained for a cubic sample of
size $N$ so that all the $P_{\xi}$ are equal. For the first
two classes of BC the $P-$ dependence appears in the Jacobi
$\theta _{3}-$ function resulting from summation over $k$ in the order index
of the Bessel function $\theta _{3}\left( \exp \left( -\pi P/n\right)
\right) $ $=\left( 1+2\ \sum_{k=1}^{\infty }\exp \left( -\pi k^{2}P/n\right)
\right) .$ The remaining sum over $n$ in (\ref{4a}) can be dealt with the
help of the known properties of the $\theta _{3}-$ function
and leads to a convergent series. In agreement with its physical
meaning the condition $P\thickapprox 1$ defines a crossover temperature $T^{\ast }\sim JS/N^{2}$ separating the exponential regime ($%
P<1$) from power law $T-$dependence ($P>1$). In the former case the low-$T$ polynomial expansion of the $I_{0}\left(
2\beta nJS\right) $ (the Bloch series) is completely canceled by the $P$%
-expansion of the finite size terms, $I_{2kN}\left( 2\beta nJS\right) $ with
$k\neq 0,$ leading to an exponential behavior $\Delta \mu ^{F}\sim \exp
\left( -\beta JS\left( \pi /N\right) ^{2}\right) +...$\ .
When considering shape anisotropy we assume that $P_{\xi} > 1 $, a case more common in experiments, when $T>T^{\ast }$. Then cancellation is
incomplete and the resulting expression can be cast into a form similar to the standard
low temperature Bloch series where each term has as a prefactor a certain
function of $P.$ These functions are calculated analytically as a
large $P$ expansion converging to the bulk limit ($P\rightarrow \infty $).

\begin{figure}[tbph]
\centering\includegraphics[height=6.2cm,width=7.6cm]{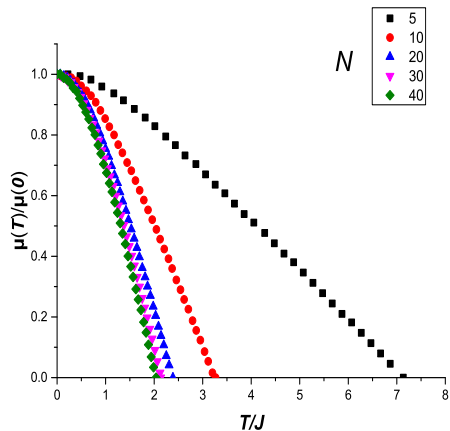} %\epsfbox{fig1.eps}
\caption{Temperature and linear size ($N$, see inset) dependence of the magnetization $\mu\left( T\right)$ for $S = 1/2$ cubic CB cluster $V = N\times N\times N$. Magnetization is higher than for the bulk material and converges to the Bloch law in the macroscopic limit, $N \rightarrow \infty $.}
\label{fig1}
\end{figure}

Below only the leading terms of the final
expressions are shown to emphasize both the contrasting behavior of the free standing and
embedded NP ($N_{z}>N_{y}>N_{x}$) and its difference from Eq.  (\ref{2a}):
\begin{eqnarray}
&& \Delta \mu ^{FB}\simeq BT^{\ 3/2}+C^{FB}\times \left( T/NJ\right) \label{FEB}\\
&& +\frac{1}{8S}\left( \frac{T}{\pi JS}\right)
^{3/2}\left( \frac{1}{\sqrt{P_{x}}}+\frac{1}{\sqrt{P_{y}}}+\frac{1}{\sqrt{%
P_{z}}}\right) \ln P_{x} ,\notag
\end{eqnarray}
\begin{eqnarray}
&& \Delta \mu ^{EB}\simeq BT^{\ 3/2}+C^{EB}\times \left( T/NJ\right) \notag\\
&& -\frac{1}{8S}\left( \frac{T}{\pi JS}\right)
\left( \frac{1}{N_{x}}+\frac{1}{N_{y}}+\frac{1}{N_{z}}\right) \ln \left(
TN_{x}^{2}/\pi JS\right). \notag
\end{eqnarray}
Here $B=\zeta \left( 3/2\right) S^{-5/2}\left( 4\pi J\right) ^{-3/2}$ is
the standard Bloch coefficient of the bulk material and  $P=P^{FB}$ henceforth. Explicit form of the $ C -$ coefficients is used in the calculations, however it is not essential for understanding the main trends. The last terms (the main finite size terms) of the two expressions in Eq.(\ref{FEB}) only differ in sign. We see that, relative to the the bulk material, magnetization is decreased (increased) for the free (embedded) sample. Unlike Eq.(\ref{2a}), this result provides a transparent representation of the physical reasons for such behavior. Note that size dependence is dominated by the smallest linear extent of the NP, e.g. nanoplates of the same thickness or nanorods of the same diameter should generally have close values of $\mu\left( T\right)$ although their volume can differ significantly. Suppression of magnetization in free particles, known from many numerical simulation and experimental studies, is due to enhanced fluctuations of the surface spins; respectively, a way to increase magnetization is embedding (or coating) the magnetic particle into a polarizable medium which couples to the surface spins. This effect is manifestly nonlinear and is strongly enhanced for smaller NP, as illustrated in Fig.~\ref{fig1}.

\begin{figure}[tbph]
\begin{center}
\includegraphics[height=5.9cm,width=7.6cm]{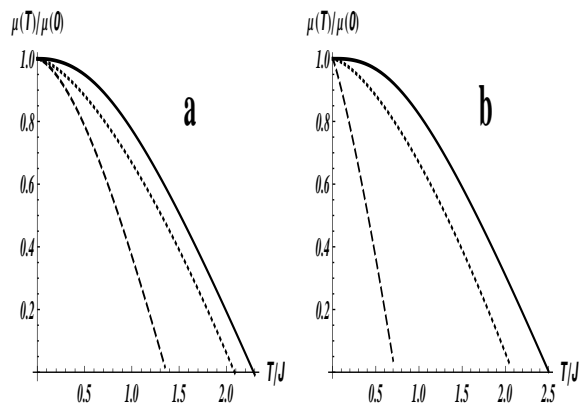} %\epsfbox{fig2.eps}
\end{center}
\caption{Magnetization curves of the two $S = 1/2$ clusters, cubic (a) $N_{x}\times N_{y}\times N_{z}= 10\times 10\times 10 $ and elongated (b) $N_{x}\times N_{y}\times N_{z}= 5\times 5\times 40 $, with the same number of spins, Eq. (8): free standing (dashed), embedded EB2 (continuous) and bulk Bloch (dotted).}
\label{fig2}
\end{figure}

For a free standing particle our results agree with the numerical diagonalization studies of \cite{Hend,Hendriksen}. The contrasting behavior of the two classes of BC is enhanced further by the shape anisotropy as shown in Fig.~\ref{fig2}. It should be kept in mind that, being based on magnon gas approximation, our description refers to the low temperature region $T\lesssim 0.5T_{C} $ and near $T_{C} $ the curves should be viewed as an extrapolation.
It is instructive to compare with PB conditions considered in an earlier publication which lead to a different functional dependence, converging much faster to the bulk limit (dotted curve in Fig.~\ref{fig2}). Moreover, unlike Eq.(\ref{FEB}), this dependence is qualitatively modified depending on the shape of the particle, as can be seen from the two examples below:
$\mu ^{PB}\left( T\right) /\mu \left( 0\right) \simeq
1-BT^{3/2}+T\ \left( 3.9-L/N\right) \ /4\pi NJS^{2}$ for the elongated
shape, $V=N\times N\times L,$ $L\geq N\gg 1;$ and $\mu ^{PB}\left( T\right)
/\mu \left( 0\right) \simeq 1-BT^{3/2}+T\ \left( 2.9-2\ln \left( N/L\right) %
\right) \ /4\pi LJS^{2}$ for the flattened shape.

Thus, the highest magnetization is reached when coupling to environment  ($\nu \neq 0$)  inhibits spin fluctuations on the interface, while the lowest corresponds to a free NP when surface spins can fluctuate more freely. However, we describe below another mechanism introduced by the third class of BC, which amends this picture and suggests additional possibilities to control magnetization. Indeed, applying mixed BC breaks the uniformity of coupling over the surface
of NP, this activates simultaneously both the opposite tendencies discussed
earlier (enhancement and suppression of spin fluctuations) resulting in a
totally different behavior. This is most convincingly demonstrated by the
conclusion that the lowest magnetization is actually achieved not for a free
standing NP but in this class of BC. The effect is formally contained in the other kind of Jacobi $\theta$ - function, $\theta _{4}\left( \exp \left( -\pi P/n\right) \right) $ which appears in our low - $T$ expansion.

\begin{figure}[tbph]
\begin{center}
\includegraphics[height=5.9cm,width=7.6cm]{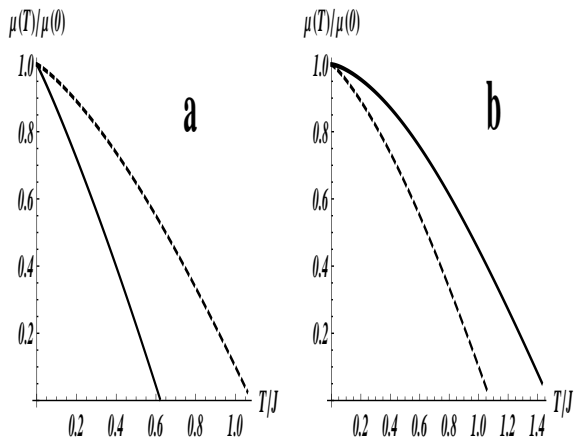} %\epsfbox{fig3.eps}
\end{center}
\caption{Magnetization curves of the same sample with non-uniform boundary conditions (MB) (continuous) for the two cases described by the Eqs.(\ref{mb1}) and (\ref{mb2}). Respectively, (a) corresponds to a cluster $N_{x}\times N_{y}\times N_{z}= 8\times 10\times 50 $ and (b) to $N_{x}\times N_{y}\times N_{z}= 50\times 10\times 8 $, see text. Dashed line stands for the free cluster (FB), Eq.(\ref{FEB}).}
\label{fig3}
\end{figure}

The outcome of the two competing tendencies is represented by the two cases when MB conditions are applied to an initially free NP : (a)  $\nu \neq 0$  at the smallest side of the crystal, $N_{z}>N_{y}>N_{x}$, and (b) $\nu \neq 0$ at the largest side of the same crystal, $N_{x}>N_{y}>N_{z}$.
\begin{eqnarray}
&&\Delta \mu _{a}^{MB}\simeq BT^{\ 3/2}+C_{a}^{MB}\times \left( T/NJ\right)\label{mb1}\\
&&+\frac{1}{4S}\left( \frac{T}{\pi JS}%
\right) \left( \frac{1}{N_{x}}+\frac{1}{N_{y}}\right) \ln \left(
TN_{z}^{2}/\pi JS\right) ,\notag
\end{eqnarray}
\begin{eqnarray}
&&\Delta \mu _{b}^{MB}\simeq BT^{\ 3/2}+C_{b}^{MB}\times \left(
T/NJ\right)+\frac{1}{4S}\left( \frac{T}{\pi JS} \right) \notag \\
&& \times\left( \frac{1}{N_{x}}\ln \left( TN_{y}^{2}/\pi JS\right) +\frac{1}{
N_{y}}\ln \left( TN_{x}^{2}/\pi JS\right) \right).\label{mb2}
\end{eqnarray}
Fig. \ref{fig3} illustrates the relation between the respective quantities $\Delta \mu _{a}^{MB}>\Delta \mu ^{FB}>\Delta \mu _{b}^{MB}$ and demonstrates the non-trivial effect of the boundary conditions and shape anisotropy on magnetization.
The unexpected large decrease of magnetization of the free NP can be traced to the modification of the magnon excitations (eigenvalues and eigenfunctions).

The above analysis of the different classes of BC allows to formulate the following generic form of the Bloch law extended to nanomagnets:
\begin{equation}
\mu \left( T\right) /\mu \left( 0\right) =1-BT^{3/2}-FT\ln T-CT,  \label{8a}
\end{equation}%
where $C$ and $F$ are size, shape and BC dependent parameters which can take
not only positive, but also  negative values ($T$ is measured in units of $J$
for the bulk material). It is qualitatively different from the standard
formula in Eq. (\ref{2a}) and leads to significantly different physical
conclusions as shown below.

In Fig. \ref{fig4} the results of quantum Monte-Carlo (QMC) simulations \cite{huang06} of a free standing cluster with $V=2123$ atoms, $S=1,$ zero
anisotropy constant $K$ and FCC crystal structure are compared to the microscopic
approach.
\begin{figure}[tbph]
\begin{center}
\includegraphics[height=6.2cm,width=7.6cm]{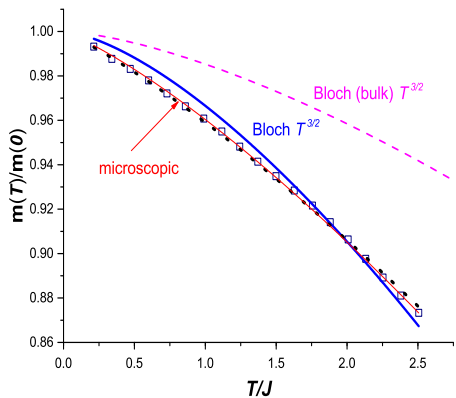} %\epsfbox{fig4.eps}
\end{center}
\caption{QMC simulations of the $V = 2123, S = 1$ free standing cluster \cite{huang06} (squares), Bloch $T 3/2$ law best fit (blue continuous), Bloch law for the bulk material (dashed magenta line), microscopic expression with FB (red continuous) and best fit with the generalized Bloch law Eq. (10) (black dotted). Best fit with the phenomenological expression Eq. (3) is not shown, see text.}
\label{fig4}
\end{figure}
The coefficients in Eq.(\ref{8a}) should be replaced by their values
corresponding to the FCC structure (e.g. $B$ is $4$ times smaller than for
the simple cubic structure etc.) and temperature is measured in units of $J$
so that we get for the bulk value $B=0.0147.$ The best fit of the cluster
magnetization by the standard Bloch law (thick continuous blue) results in a value almost three times larger, $B=
B_{tot}=0.034 $, than the bulk (dashed thin magenta). According to the current interpretation, e.g. \cite{gubin}, the Bloch coefficient $B$ is an average of the core,  $B_{core}$, and surface, $B_{surface}$, contributions. Thus, respective quantities obtained in \cite{huang06} are $B_{core}=0.0245$ and $B_{surface}=0.0468$. Consequently, the large value of $B$ would imply an overall softening of the magnon spectrum and a large decrease of the effective coupling constant $J$ to almost half of its initial (bulk) value, clearly demonstrating the inconsistency of the approach. An even larger softening should be inferred from the
phenomenological expression (\ref{2a}) which gives $\gamma =0.04$. The respective curve is not shown in the
figure to avoid confusion with the present calculation and the MC simulation data. The numerical closeness of the results demonstrates the validity
of the microscopic approach because in this case the exchange coupling constant remains unchanged, as it should, and $B=0.0147$, the same as in the bulk limit, while the size and surface effects
are explicitly taken into account. These are represented by the $T\ln T$ and
linear $T$ terms with $F=0.002$ and $C=0.025$ as the corresponding
prefactors in Eq.(\ref{8a}) for $\Delta \mu ^{FB}$  (thin red line).
If we now use the generalized form of the Bloch law in Eq. (\ref{8a}) and determine the respective coefficients by fitting the Monte-Caro simulation data of Ref. \cite{huang06} (squares), we then obtain (dotted line): B = 0.0178, F = 0.002 and C = 0.021. I.e., the value of $J$ estimated on the basis of Eq. (\ref{8a}) is only a few percents off the exact result. This analysis clearly demonstrates that a simple $T^{3/2}$ Bloch law may indeed appear to represent the behavior reported in some experimental works, however the values of the Bloch coefficient extracted in this way disagree with the microscopic physics of the system and grossly overestimate magnetic softening.

\section{Conclusions}

The microscopic description of the temperature dependent magnetization $\mu\left( T\right)$ of ferromagnetic nanostructures with relatively weak magnetic anisotropy leads to the generalized form of the Bloch law given by Eq.(\ref{8a}), different from the one presently used for the analysis of experimental and numerical simulation data, Eq.(\ref{2a}). We have demonstrated with a specific example that the latter is misleading, despite of its good performance as a data fitting formula. In particular, it largely overestimates magnetic softening in nanostrutures and the true softening is better captured by the proposed form of the modified Bloch law. Magnetization consists of two contributions: the Bloch term for the bulk material, and the size-dependent part including effects of geometric shape and coupling to surrounding medium. Such coupling can strongly modify the properties of a nanostructure through a mechanism mediated by collective excitations. Boundary conditions can cause a large variation of magnetic polarization. For instance, embedding a free unpolarized NP in a polarizable medium may induce a magnetic moment which can exceed even the bulk value, an effect strongly enhanced by shape anisotropy. For an asymmetric shape one can increase or even decrease the magnetization of a free standing NP depending on the contact area, suggesting an interesting possibility of controlling magnetization. It turns out that in the case of non-uniform coupling suppression of spin fluctuations on a part of the surface can be prevailed by the enhanced fluctuations of the other spins of the sample. Aside from possible experimental realizations of these effects, the approach may be useful for the description of other kinds of excitations in nanocrystals.

\acknowledgments
One of the authors (SC) gratefully acknowledges stimulating discussions with
Dr.\ D.V. Anghel. The work has been financially supported by CNCSIS-UEFISCDI
(project IDEI 114/2011) and by ANCS (project PN 09370102).

\end{document}